\documentstyle[12pt,aaspp4]{article}
\begin{document}
\parindent=1.0cm

\title{NEAR-INFRARED ADAPTIVE OPTICS IMAGING OF THE CENTRAL REGIONS OF NEARBY 
Sc GALAXIES: I. M33}

\author{T. J. Davidge \altaffilmark{1}}

\affil{Canadian Gemini Office, Herzberg Institute of Astrophysics,
\\National Research Council of Canada, 5071 W. Saanich Road,\\Victoria
B.C. Canada V8X 4M6\\ {\it email:tim.davidge@hia.nrc.ca}}

\altaffiltext{1}{Visiting Astronomer, Canada-France-Hawaii Telescope, which is 
operated by the National Research Council of Canada, the Centre National de la 
Recherche Scientifique, and the University of Hawaii}

\begin{abstract}

	Near-infrared images obtained with the Canada-France-Hawaii Telescope 
(CFHT) Adaptive Optics Bonnette (AOB) are used to investigate the stellar 
content within 18 arcsec of the center of the Local Group spiral galaxy M33. 
AGB stars with near-infrared spectral-energy distributions similar to 
those of giants in the solar neighborhood and 
Baade's Window are detected over most of the field. Neither the peak brightness 
nor the color of the AGB sequence on the $(K, J-K)$ color-magnitude diagram 
changes with distance from the galaxy center. The 
bolometric luminosity function (LF) of these stars has 
a discontinuity near M$_{bol} = -5.25$, and comparisons with evolutionary 
tracks suggest that most of the AGB stars formed in a burst of 
star formation $1 - 3$ Gyr in the past, indicating that the star 
formation rate near the center of M33 has varied significantly during the 
past few Gyr.

	The images are also used to investigate the integrated near-infrared 
photometric properties of the nucleus and the central light concentration. 
The nucleus is bluer than the central light concentration, 
in agreement with previous studies at visible wavelengths. The 
near-infrared photometric properties of the nucleus are reminiscent of 
relatively young clusters in the Magellanic Clouds, while the photometric 
properties of the central light concentration are similar to those 
of globular clusters. The CO index of the central light concentration 
0.5 arcsec from the galaxy center is 0.05, which 
corresponds to [Fe/H] $\sim -1.2$ for simple stellar systems. 
Hence, the central light concentration could not have 
formed from the chemically-enriched material that dominates the 
present-day inner disk of M33.

\vspace{0.3cm}

\end{abstract}

\section{INTRODUCTION}

	The central regions of spiral galaxies do not form and evolve in 
isolation. Observations indicate that tidal interactions (e.g. Barnes \& 
Hernquist 1992 and references therein) can trigger the transport of material 
from disks to the central regions of galaxies, while bar instabilities (e.g. 
Ho, Filippenko, \& Sargent 1997 and references therein), which may 
dissolve to form bulge-like structures (e.g. Norman, Sellwood, \& Hasan 1996), 
can transport material to small galactic radii. Moreover, 
if galaxies form in a hierarchal manner, as is required to 
match the integrated light density of the Universe as a function of redshift 
(e.g. Madau, Pozzetti, \& Dickinson 1998), then the bulges of spiral galaxies 
will almost certainly accumulate stars spanning a range of chemical 
compositions and ages from other systems.

	Nearby spiral galaxies are important laboratories for studying bulge 
evolution. The Galaxy is of particular interest as the age of 
the bulge can be measured directly from the brightness of the main sequence 
turn-off, and the presence of metal-rich bulge globular clusters 
(Minniti 1995) with halo-like ages (Ortolani et al. 1995, Fullton et al. 
1995) indicates that at least some portions of the Galactic bulge formed 
during early epochs. However, the Galactic bulge may not consist 
exclusively of old stars. Andredakis, Peletier, \& Balcells (1995) find a 
smooth change in the central structural characteristics of spiral galaxies with 
Hubble type, suggesting that interactions with the disk have progressively 
greater impact on the central properties of galaxies with later morphological 
types; hence, it might be anticipated that the contribution of younger stars 
to the light from the central regions of galaxies will increase 
towards later morphological types. The structural characteristics of bulges 
may also change with morphological type. Phillips et al. (1996) find that 
the central surface brightness profiles of spiral galaxies undergo 
a change near type Sc (Phillips et al. 1996), in the sense that the 
bulges of systems earlier than type Sc appear to be 
related to early-type galaxies, while the bulges of systems later than type 
Sc appear to be structurally similar to nucleated dwarf galaxies. 
These data suggest that the central light concentrations in Sc galaxies may 
be transitional objects, with characteristics bridging 
earlier and later morphological types. Thus, studies of Sc galaxies may 
provide important insight into the physical processes that determine bulge 
evolution. 

	This is the first in a series of papers dealing with 
high angular resolution near-infrared observations 
of the central regions of nearby Sc galaxies 
obtained with the CFHT AOB. The near-infrared 
is of interest for studies of this nature since 
the light at these wavelengths originates mainly from old and 
intermediate-age populations, which formed when the structural characteristics 
of the central regions of the host systems were imprinted. The 
contribution of line and continuum emission to the total light output is 
also smaller in the near-infrared than at visible wavelengths, thereby 
simplifying efforts to study stellar content. Finally, angular resolutions 
approaching the diffraction limit of the CFHT can be obtained with the AOB at 
these wavelengths.

	With a distance modulus $\mu_0 = 24.5 \pm 0.2$ (van den Bergh 1991), 
M33 is a logical starting point for a study of the central regions of Sc 
galaxies. The surface brightness profile of M33 departs from 
an exponential near the galaxy center (e.g. Kent 1987, Bothun 
1992, Regan \& Vogel 1994), indicating that the central light concentration is 
structurally distinct from the disk. Bothun (1992) suggests that the central 
light concentration may be the inner extension of the halo, although this 
conclusion is based in part on data having relatively coarse angular 
resolution. In fact, images with subarcsec resolution reveal that
the central light concentration contains a blue nucleus (Kormendy \& McClure 
1993, Mighell \& Rich 1995, Lauer et al 1998). The nucleus is very compact, and 
at ultraviolet wavelengths 50\% of the nuclear light falls within the central 
0.14 arcsec (Dubus, Long, \& Charles 1999). The central light concentration 
is slightly flattened (Lauer et al. 1998), and has an extremely 
high surface density, such that the innermost regions have 
likely experienced core collapse (Kormendy \& McClure 1993, Lauer et al. 1998).

	There are indications that M33 may harbour an active nucleus. Rubin \& 
Ford (1986) detected [NII] emission in the nuclear spectrum, while 
Matthews et al. (1999) find what appears to be a jet originating from the 
nucleus. The nucleus also contains a bright x-ray source (Long et al. 1981) 
which, although having a spectrum suggestive of low-level AGN behaviour, 
may instead be an x-ray binary (Trinchieri, Fabbiano \& 
Peres 1988). The nucleus of M33 does not contain a 
super-massive black hole (Kormendy \& McLure 1993), and the absence of such an 
object means that the dynamical evolution of the central regions will be very 
different from systems like M31 or the Milky-Way (Lauer et al. 1998). Finally, 
Regan \& Vogel (1994) find evidence of a weak bar, which may have 
played a significant role in the recent evolution of the nucleus. 

	Given that star formation is occuring throughout the disk of M33 
it is perhaps not surprising that the central regions of this galaxy contain 
young and intermediate-age components. Spectroscopic investigations by 
O'Connell (1983) and Schmidt, Bica, \& Alloin (1990) indicate that a 
substantial fraction of the visible light within a few arcsec of the galaxy 
center originates from stars with ages $\leq 1$ Gyr, and that the most 
chemically evolved stars in this region have solar metallicities. Gordon et al. 
(1999) examine spectra covering a broad range of wavelengths and conclude 
that star formation in the nucleus has not been continuous, and that 
the most recent burst of star formation occured 70 -- 75 Myr in the past.
Gordon et al. also conclude that the central regions of M33 contain 
dust that follows an extinction curve like that in the Milky-Way.

	The presence of intermediate age stars near the 
center of M33 have been confirmed by imaging studies. 
Minniti, Olszewski, \& Rieke (1993) used near-infrared images to detect stars 
as luminous as M$_{bol} \sim -6$ near the galaxy center, although subsequent 
observations by McLean \& Liu (1996), who studied the near-infrared properties 
of stars with distances in excess of 45 arcsec from the nucleus to avoid 
crowding, suggest that the brightest objects in the Minniti et al. sample may 
be blends of fainter stars. McLean \& Liu (1996) found a 
discontinuity in the luminosity function (LF) near M$_{bol} \sim -5.3$, which 
was interpreted as the AGB-tip of a population with an age of $1 - 2$ Gyr.

	The highest angular resolution investigation of resolved stars near the 
center of M33 was conducted by Mighell \& Rich (1995) using archival $V$ and 
$I$ WFPC1 images. These data revealed a large population of blue main sequence 
stars, which are less centrally concentrated than the old and intermediate-age 
components. Giant branch stars were also resolved, and the width of the RGB 
suggests that the inner regions of M33 experienced a 
chemical enrichment history that is very different from the outer 
halo field studied by Mould \& Kristian (1986).

	In summary, the existing data indicate that the central regions of M33 
contain a mixture of stellar populations, spanning a range of ages and 
metallicities. Crowding has presented a significant obstacle for efforts to 
probe the regions in and around the central light concentration, and in the 
current study near-infrared images with $\sim 0.3$ arcsec angular resolution 
are used to investigate the stellar content within 18 arcsec 
of the galaxy center. Details of the observations 
and the methods used to reduce the data are described in \S 2, while the 
near-infrared photometric properties of resolved stars in the circumnuclear 
region are discussed in \S 3. The stellar content of the central light 
concentration is investigated in \S 4, and a discussion and summary of the 
results follows in \S 5.

\section{OBSERVATIONS \& REDUCTIONS}

	The data were obtained during the night of UT Sept 6/7 1998 with the 
KIR imager, which contains a $1024 \times 1024$ Hg:Cd:Te array with 0.034 
arcsec pixels, and CFHT AO system (Rigaut et al. 1998). Images were recorded 
through $J, H, K, Br\gamma,$ CO, and $2.2\mu$m continuum filters with the 
galaxy nucleus centered on the detector; the nucleus also served as the 
reference source for AO compensation. Four groups of five 
exposures were recorded in each filter, with the telescope offset after each 
group to form a $0.5 \times 0.5$ arcsec dither pattern. 
The total integration time was 20 minutes per filter, except for 
CO, where the total exposure time was 40 minutes. The FWHM of 
individual stellar images is 0.34 arcsec in all filters. 

	Calibration frames were constructed using the procedures described by 
Davidge \& Courteau (1999). The raw data were corrected for detector dark 
count, flat-field variations, DC sky level, and thermal emission from objects 
along the optical path. The results were aligned on a filter-by-filter basis to 
correct for the offsets introduced at the telescope, and then median-combined 
to reject cosmic rays and bad pixels. The final $K$ image is shown in Figure 1. 
The $J$ and $H$ images are very similar to the $K$ image, and hence are not 
shown here.

\section{THE NATURE OF RESOLVED STARS NEAR THE CENTER OF M33}

\subsection{Photometric measurements}

	Stars are resolved over much of the KIR field, and the brightnesses of 
these were measured using the PSF-fitting routine ALLSTAR (Stetson \& Harris 
1988). Photometric measurements are complicated by the surface-brightness 
distribution of M33, which changes significantly on sub-arcsec angular 
scales near the galaxy center, thereby making it difficult to measure 
local sky levels. This background structure was removed using the iterative 
technique described by Davidge, Le F\`{e}vre, \& Clark (1991).

	A single PSF was constructed for each filter using routines in the 
DAOPHOT (Stetson 1987) photometry package. While isoplanicity causes the PSF to 
vary with distance from the galaxy center, which served as the 
reference source for AO compensation, the star-subtracted images produced by 
ALLSTAR do not show obvious systematic residuals indicative of radial PSF 
variations. Moreover, observations of globular clusters have shown that 
during typical seeing conditions PSF variations over the KIR field 
do not affect photometric measurements by more than a few percent 
(Davidge 1999, Davidge \& Courteau 1999). 

	Standard stars from Casali \& Hawarden (1992) and Elias et al. 
(1982) were observed throughout the 4 night observing run, and 
photometric zeropoints were computed from the resulting images. 
The uncertainty in the zeropoints is typically $\pm 0.03$ mag.

	The brightnesses and colors of stars with $K \leq 17.5$ are listed 
in Table 1. The $x$ and $y$ entries in this Table are offsets, in pixels, 
from the center of M33 in the reference frame defined by Figure 1, which is 
1000 pixels on a side. Positive offsets are to the West and North.

\subsection{Photometric properties}

	The $J, H,$ and $K$ LFs of the circumnuclear region are shown in Figure 
2. The LFs follow power-laws at the bright end, and the drop in number counts 
at the faint end indicates that incompleteness becomes significant at $J = H = 
20$, and $K = 19$. The $(K, H-K)$ and $(K, J-K)$ CMDs, shown in Figure 3, 
contain a plume of stars extending from $K = 18$ to $K = 16$. 
The scatter in the two CMDs is similar, amounting to $\sigma_{HK} = 
\pm 0.12$ mag and $\sigma_{JK} = \pm 0.13$ mag
between $K = 17.5$ and 18. $J-K$ is more sensitive to 
variations in age and metallicity than $H-K$, so this similarity in scatter 
suggests that observational errors, rather than the intrinsic 
properties of stars, dominant the dispersion along the color axes.

	The statistical uncertainties in the photometric measurements have 
also been estimated with artificial star experiments. 
Artificial stars with colors and brightnesses that follow
the locus of points on the CMDs were added to the $J, H,$ and $K$ images 
using the DAOPHOT task ADDSTAR, and the brightnesses of the added stars were 
then measured with ALLSTAR. The standard deviations of the difference between 
the measured and actual brightnesses were computed, and the results were 
used to generate the error bars plotted in Figure 3. The 
uncertainties estimated in this manner are in reasonable agreement 
with the scatter on the CMDs, providing further evidence that observational 
errors are the dominant source of scatter.

	At the distance of M33 the RGB-tip for an old solar metallicity 
population would be expected to occur near $K \sim 18.5$, and a significant 
number of stars brighter than this have been detected 
near the center of M33. Crowding is a major 
concern in high-density environments, as spurious bright objects are 
produced when two or more moderately faint stars fall within 
the same resolution element. Therefore, the possibility that the sources 
with $K \leq 18.0$ may not be individual stars must be investigated. 

	If the circumnuclear region consists exclusively of an old population 
like that in metal-rich Galactic globular clusters then objects significantly 
brighter than $K \sim 18.5$ will be the result of two (or more) giants 
within $\sim 0.5$ mag of the RGB-tip occuring in the same resolution element. 
NGC6316 is a moderately metal-rich globular cluster that was observed by 
Davidge et al. (1992) in $K$, and the results from that study can be 
used to assess the effects of blending in the current dataset. 
Davidge et al. (1992) found 4 stars in the top 0.5 mag interval of the 
NGC6316 giant branch in $K$ over a field covering the cluster half mass 
radius; hence, NGC6316 likely contains $2 \times 4 = 8$ stars in the top 0.5 
mag of the giant branch. The surface brightness of M33 10 arcsec from the 
galaxy center is $\mu_v \sim 19.5$ mag/arcsec$^2$ (Kent 1987), so that each 
0.34 arcsec wide resolution element samples a population with M$_V = -2.4$ 
if $\mu_0 = 24.5$. Given that the integrated brightness of NGC6316 is M$_V = 
-8.6$ (Harris 1996) then, if the center of M33 is dominated by an old 
moderately metal-rich population, each KIR resolution element should 
contain $\sim 0.026$ upper giant branch stars. The probability of having two 
such stars in one resolution element is then $\sim 0.026^2 = 7 \times 10^{-4}$, 
so that there will be roughly 9 objects with $K \sim 18$ that are the result of 
blends in the current dataset. For comparison, more 
than 300 objects brighter than $K = 18$ have been detected. 
The incidence of blends plummets when K $\leq 18$; for example, a blend 
of 4 RGB-tip stars is required to produce an object with $K =17$, and the 
probability of this occuring is $5 \times 10^{-7}$, so that there would be 
0.006 such objects per KIR field. Thus, the central regions of M33 contain 
stars that are brighter than would be expected from an old moderately 
metal-rich population.

	Neither the peak brightness nor mean color of the CMD ridgeline 
varies across the KIR field. This is demonstrated in Figure 4, 
where the $(K, J-K)$ CMDs of stars in 3 annuli, centered on the M33 nucleus, 
are plotted. The absence of a correlation between peak stellar brightness 
and distance from the galaxy center provides further evidence that the 
brightest sources are individual stars, and not blends.

	The $(M_K, (J-K)_0)$ CMDs of stars near the center of M33 are compared 
with Large and Small Magellanic Cloud red supergiants (RSGs) and long period 
variables (LPVs) from the studies of Elias, Frogel, \& Humphreys (1985) and 
Wood, Bessell, \& Fox (1983) in Figure 5. The RSGs define a vertical plume on 
the CMD, and the corresponding portion of the M33 CMD is unoccupied. However, 
the M33 data overlap with the Magellanic Cloud LPV observations, although the 
latter dataset contains stars that are almost 2 mag brighter than the brightest 
M33 stars. Hence, the recent star forming histories of the inner disk of 
M33 and the Magellanic Clouds have been very different.

	The sequence defined by M giants in Baade's Window (BW), which is also 
plotted in Figures 3 and 4, falls to the right of the M33 data, 
ostensibly suggesting that stars surrounding the central light concentration 
may be more metal-poor than those in BW. However, M33 contains stars that are 
intrinsically brighter than those in BW, indicating a difference in ages, 
which will in turn affect giant branch colors. In fact, the near-infrared 
spectral-energy distributions (SEDs) of stars in M33 and BW are not 
significantly different. This is demonstrated in Figure 6, which shows 
the $(J-H, H-K)$ two-color diagram (TCD) for stars with $K \leq 18$. To reduce 
scatter on the TCD, mean colors were computed in $\pm 0.25$ mag intervals 
along the $K$ axis of the $(K, H-K)$ and $(K, J-K)$ CMDs, and 
the points plotted in Figure 6 are based on these means. The M33 points lie 
between the solar neighborhood and BW giant sequences 
in Figure 6 and, with $\pm 0.04$ mag uncertainties in the color calibrations, 
the M33 data are not significantly different from the BW sequence.

\subsection{Comparisons with models and the center of the Milky-Way}

	Davidge (1998b) used the solar metallicity models of 
Bertelli et al. (1994) to generate near-infrared AGB sequences. These sequences 
are compared with the M33 observations in Figure 5, and it is evident that most 
of the stars detected near the center of M33 have an age between 1 and 10 Gyr, 
while there are few stars with ages between 0.1 and 1 Gyr.

	The colors of theoretical sequences on CMDs are sensitive to 
uncertain quantities such as the mixing length and the relation between color 
and effective temperature used to tranform stellar structure models onto the 
observational plane. The bolometric LF provides a more direct 
means of investigating the ages of stellar populations. Bolometric 
luminosities were computed for stars near the center of M33 using the 
relations between the bolometric correction, BC$_K$, and $J-K$ for solar 
neighborhood and BW giants from Figure 1b of Frogel \& Whitford (1987). The 
resulting LFs, which are compared in the upper panel of Figure 7, are almost 
identical, and follow a power-law from M$_{bol} \sim -5$ to the onset of 
incompleteness at M$_{bol} \sim -3.5$. The power-law exponent, derived from a 
least squares fit to the 4 LF bins with M$_{bol}$ between --5 and --3.5, is 
$0.528 \pm 0.036$.

	Discontinuities in the LF near the bright end are one indication that 
the recent star forming history near the center of M33 has not been continuous. 
The LF in Figure 7 appears to depart from the trend infered from moderately 
faint bins when M$_{bol} \leq -5.25$, which is the AGB-tip luminosity for a 
population with an age log(t) = 9.2 $\pm 0.2$, where $t$ is in Gyr, 
according to the solar metallicity Bertelli et al. (1994) models. 
The significance of this departure from the least squares 
fit shown in Figure 7 is investigated in Figure 8, where the 
differences between the observed and predicted numbers of stars in each LF bin 
and the ratio of these differences to the estimated uncertainties, which 
provides a direct measure of statistical significance, are plotted. It is 
evident from the lower panel of Figure 8 that the difference between the 
observations and the least squares fit is significant at the $2.5\sigma$ level 
or higher when M$_{bol} \leq -5.0$, indicating that a significant break in the 
LF occurs when M$_{bol} = -5.25 \pm 0.25$. The presence of such a break 
indicates that the majority of AGB stars with M$_{bol} \geq -5.25$ have ages 
1 -- 3 Gyr, and that star forming activity declined $\sim 1$ Gyr in the past; 
consequently, the star formation rate near the center of M33 has not been 
constant during the last few Gyr. 

	Minniti et al. (1993) note that the LF of stars near the center of 
M33 is similar to that of the Galactic Center, 
and in the lower panel of Figure 7 the bolometric 
LFs for stars near the centers of M33 and the Galaxy are compared. The 
Galactic LF was constructed from the observations of the field surrounding SgrA 
obtained by Davidge (1998a). It is apparent that the M33 and Galactic LFs have 
very similar power-law exponents in the interval $-3.5 \geq$ M$_{bol} \geq -5$.

\section{PROPERTIES OF THE CENTRAL LIGHT CONCENTRATION}

\subsection{Photometric gradients}

	Kormendy \& McClure (1993) and Lauer et al. (1998) found that the 
nucleus of M33 is bluer than the surroundings 
at visible wavelengths. To determine if this is also the case 
in the infrared, mean $J-K$, $H-K$, CO, and 
Br$\gamma$ values were computed in 0.1 arcsec wide annuli centered on the M33 
nucleus, and the results are shown in Figure 9. 

	The background sky level, which is of critical importance for measuring 
colors in low surface-brightness regions, can not be 
reliably measured from the KIR data given the extended nature of M33. Although 
a blank sky field 1 degree from the galaxy center was observed immediately 
following the M33 observations, the sky brightness at infrared 
wavelengths can change significantly over time scales of a few minutes and 
angular offsets of a few arcmin. Rather than rely on potentially uncertain sky 
measurements from the offset field, background sky levels 
for each filter were computed by assuming that the 
surface brightness profile near the center of M33 follows an 
r$^{1/4}$ profile, as indicated by observations at visible wavelengths (e.g 
Bothun 1992, Regan \& Vogel 1994). This profile was fit to the data 
between 1 and 3.5 arcsec from the galaxy center to satisfy the conflicting 
requirements of avoiding contamination from the bright nucleus while 
maintaining an acceptably high S/N ratio for surface brightness measurements. 
It should be emphasized that the color measurements have been restricted to the 
central 1.5 arcsec of M33, where the surface brightness is high and 
photometric measurements are insensitive to the adopted sky levels.

	The $J-K$ and $H-K$ profiles in Figure 9 indicate that the 
nucleus is bluer than the surroundings, in qualitative agreement with the 
Kormendy \& McClure (1993) and Lauer et al. (1998) measurements. The 
CO index strengthens with decreasing radius, reaching 0.24 within 0.1 arcsec of 
the nucleus, indicating that a significant population of bright cool stars is 
present near the center of M33. Gordon et al. (1999) used $2\mu$m spectra 
obtained during 1 arcsec seeing to estimate the CO index within 0.6 arcsec 
of the center of M33, and found that CO = 0.21. To determine if 
the current data produce a similar result, the CO and continuum images were 
convolved with a Gaussian to simulate 1 arcsec FWHM seeing. The CO index 
within a 1.2 arcsec aperture was then measured, with the result that CO 
= 0.21. Thus, the current CO measurements are consistent with 
the Gordon et al. (1999) result.

	The radial Br$\gamma$ profile in Figure 9 is suggestive 
of strengthening absorption towards smaller radii in the pass band of the 
Br$\gamma$ filter. The nucleus of M33 has a very early spectral type (van den 
Bergh 1976) with strong H$\alpha$ absorption (e.g. Rubin 
\& Ford 1986), and it is tempting to interpret the radial Br$\gamma$ profile in 
the context of very strong nuclear Br$\gamma$ absorption. However, the 
near-infrared photometric properties of the M33 nucleus indicate that 
bright red stars are present (see above), and it is not clear if strong 
Br$\gamma$ absorption would result in such a situation. Other species, 
such as CN, which is prominent in Carbon star spectra at these wavelengths 
(e.g. Davidge 1990), could dominate the Br$\gamma$ filter pass band near the 
nucleus. High angular resolution spectroscopy covering 
Br$\gamma$ would be of interest to determine the cause of the radial 
variation in the Br$\gamma$ index.

\subsection{The stellar content of the central light concentration}

	$(J-H, H-K)$ and $(CO, J-K)$ TCDs provide insight into the stellar 
content of the nucleus and the surrounding central light concentration. 
These TCDs for the central regions of M33 are shown in 
in Figure 10, where the plotted points are the normals calculated in \S 4.1.

	Frogel (1985) measured the central colors of nearby Sc galaxies through 
a 6.6 arcsec diameter aperture. These measurements are plotted in Figure 10, 
and it is evident that they overlap with the M33 datapoints that sample the 
outer parts of the central light concentration. The agreement is best on the 
$(CO, J-K)$ TCD, where the Frogel (1985) data continue the sequence defined by 
the M33 observations. This comparison indicates that the colors and CO 
indices measured for M33 are consistent with those seen in other systems.

	Also plotted in Figure 10 are datapoints for M31 globular clusters 
(Frogel, Persson, \& Cohen 1980) and open clusters in the 
Magellanic Clouds (Persson et al. 1983). The cluster and M33 data 
occupy similar regions of the TCDs, indicating that the near-infrared 
light from the central few arcsec of M33 is mainly stellar in origin. 

	The $(CO, J-K)$ TCD is of particular interest as 
Magellanic Cloud clusters of Searle, Wilkinson, \& Bagnuolo (1980; hereafter 
SWB) types 1 and 2 define a distinct sequence on this diagram, 
in the sense of having relatively strong CO absorption coupled with moderately 
blue $J-K$ colors. The photometric properties of SWB type 1 and 2 clusters 
can be understood in the context of age: they are among the youngest in the 
Magellanic Clouds (e.g. Hodge 1983) and the strong CO indices originate 
from very luminous AGB-tip stars (Frogel, Mould, \& Blanco 1990), while the 
blue broad-band colors are a consequence of relatively hot giant branch 
temperatures, combined with a bright, blue, main sequence turn off. 

	The nucleus of M33 has $J-K = 0.7$ and CO = 0.25, whereas at larger 
radii CO = 0.05 and $J-K \geq 0.7$. It is evident from the 
lower panel of Figure 10 that the photometric characteristics 
of the nucleus are reminiscent of SWB types 1 and 2 clusters, while 
at distances in excess of 0.4 arcsec from the nucleus, the M33 
datapoints coincide with those of M31 globular clusters 
and Magellanic Cloud clusters with later SWB classifications. Given that SWB 
types 1 and 2 clusters are much younger than M31 
globular clusters, these data thus suggest that the color 
gradient near the center of M33 is due at least in part to age effects, in the 
sense that there is a relatively young nucleus surrounded by older stars.

\subsection{ Comparison with SgrA}

	A young stellar population is present in the SgrA 
complex (Davidge et al. 1997 and references therein), and Massey 
et al. (1996) speculate that the nucleus of M33 may contain a population of 
hot stars similar to those in SgrA. It is clearly of interest to compare the 
near-infrared photometric properties of the M33 nucleus and the Galactic 
Center. To conduct such a comparison, the wide-field mosaiced images of SgrA 
and environs obtained by Davidge (1998a) 
were block-averaged to reproduce the spatial sampling obtained 
by AOB+KIR at the distance of M33, and the results were 
convolved with the PSFs constructed for the photometric analysis in \S 3. 
The background sky levels for the Galactic Center data were fixed so that the 
$2\mu$m light profile followed that plotted in Figure 1 of Saha, Bicknell, \& 
McGregor (1996).

	Following the procedure described in \S 4.1, $J-K$ and CO 
were measured in the block-averaged and seeing-convolved SgrA images 
by radially averaging these quantities in 0.1 arcsec wide 
annuli, and the results are shown in Figure 11. The data plotted in this 
figure have not been corrected for extinction and, aside from a constant offset 
due to differences between the mean extinctions of the two fields, the 
$J-K$ and CO profiles for SgrA are very different from those of the central 
regions of M33, in the sense that weaker CO absorption and larger $J-K$ colors 
are seen with decreasing radius near SgrA. While there is a 
known CO deficiency in the vicinity of SgrA* (e.g. Haller et al. 
1996, Sellgren et al. 1990), this is restricted to the central 8 arcsec of 
the Galaxy, which corresponds to 0.08 arcsec at the distance of M33. Hence, 
the depressed CO level in Figure 11, which occurs over much larger angular 
scales, likely has a different origin.

	The extinction towards the Galactic Center is not uniform, and the 
near-infrared colors of individual stars suggest that A$_K$ may reach a 
local maximum of $\sim 3.1$ mag within 40 arcsec of the Galactic Center, 
which corresponds to $\sim 0.4$ arcsec at the distance of M33, 
with A$_K = 2.8$ mag in the surrounding areas (Davidge 1998a). 
Both $J-K$ and CO would be affected by significant amounts if A$_K$ 
changed by 0.3 mag. In particular, a 0.3 mag decrease 
in A$_K$ would increase $J-K$ by 0.5 mag, and lower 
CO by 0.04 mag; hence, a radial extinction gradient 
like that proposed by Davidge (1998a) could contribute significantly to the 
$J-K$ and CO profiles in Figure 11. 

	The nuclear CO indices and $J-K$ colors of M33 and the Galaxy 
at 0.1 arcsec radius are compared in Table 2. The de-reddened colors, shown 
in the last two columns, assume that $E(B-V) = 0.3$ for M33, and A$_K = 3.1$ 
for SgrA. The errors listed in this table assume a $\pm 0.03$ mag 
uncertainty in the photometric calibration, a $\pm 0.1$ mag uncertainty 
in A$_K$ towards SgrA, and a $\pm 0.1$ mag uncertainty in $E(B-V)$ towards M33.
A solar neighborhood extinction curve has also been adopted. It is 
evident from Table 2 that the CO indices for M33 and SgrA are 
in excellent agreement, while the $J-K$ values differ at roughly the 
$2-\sigma$ level. While the difference between the $J-K$ colors is only 
marginally significant, it underlines the need to obtain integrated 
near-infrared spectra of the nucleus of M33 and the SgrA complex to provide a 
firmer basis of comparison.

\section{DISCUSSION \& SUMMARY}

	Near-infrared images with 0.3 arcsec angular resolution have been used 
to investigate the stellar content near the center of the Local Group Sc 
galaxy M33. The brightest resolved objects in the innermost regions of the 
M33 disk are luminous AGB stars, and the sources detected with the current data 
are likely the infrared-bright stars studied by Mighell \& Rich (1995). 
The brightest AGB stars in the inner disk of M33 
are almost 2 mag fainter in $K$ than the brightest AGB stars in the Magellanic 
Clouds and late-type spirals in the Sculptor group (Davidge 1998b). The 
near-infrared SEDs of the M33 stars are characteristic of moderately metal-rich 
populations, as expected given that the metallicity 
of gas near the center of M33, infered from studies of HII regions, is 
comparable to that in the solar neighborhood (Garnett et al. 1997, Shaver et 
al. 1983).

	There is a break in the AGB LF at M$_{bol} \sim -5.25$, suggesting that 
the star formation rate in the inner disk of M33 $1 - 3$ Gyr in the past was 
elevated with respect to more recent epochs. McLean \& Liu (1996) also detected 
a discontinuity at this luminosity at larger distances from the galaxy center, 
indicating that this enhanced level of star formation occured over a 
significantly larger area than the KIR field. While the absence 
of very luminous AGB stars indicates that the star formation rate 
changed $\sim 1$ Gyr in the past, the presence of blue main sequence stars 
(Mighell \& Rich 1995) indicates that recent star formation has occured in 
the innermost regions of the M33 disk.

	If the statistics of bright giants in the moderately metal-rich 
globular cluster NGC6316 are assumed to hold for the inner disk of M33 then 
a significant fraction of the sources detected in the 
current study with $K \geq 18$ (i.e. near the RGB-tip) may not be individual 
stars, but rather blends of fainter stars. However, the situation is very 
different for objects brighter than $K = 18$. For example, a blend of 4 RGB-tip 
stars would be required to produce an object with K = 17, and the chance of 
this happening in the KIR field is negligible. If the brightest 
objects detected in the current study were blends of two or more faint 
stars then the peak stellar brightness should vary with distance from the 
center of M33, and this is not seen. In fact, the AGB-tip brightness in the 
innermost regions of the M33 disk appears to be similar 
to that seen at much larger radii, as is the case in late-type spirals 
in the Sculptor Group (Davidge 1998b).

	The near-infrared colors of the central light concentration, which is 
not resolved into stars with the current data, vary with distance from the 
galaxy center, in the sense that the nucleus is bluer than the surroundings. 
This trend is consistent with observations at visible wavelengths (Kormendy \& 
McClure 1993, Lauer et al. 1998). Kormendy \& McClure (1993) and Lauer et al. 
(1998) noted that the nucleus of M33 may have experienced core collapse, and 
that the blue population may be the result of stellar mergers. However, this 
explanation for the blue population is not without problems. For example, the 
collision time scale near the center of M32 is only $3\times$ lower than in 
M33, and there is no evidence for a significant population of blue merger 
remnants near the center of the former galaxy (Lauer et al. 1998). Indeed, 
Gordon et al. (1999) reinvestigated the incidence of mergers near the center of 
M33 and concluded that Lauer et al. (1998) have likely overestimated the 
frequency of collisions.

	A recent episode of star formation near the center of M33 appears to 
offer the most reasonable explanation for the blue nucleus. Indeed, the 
near-infrared photometric characteristics of the nucleus are similar to those 
of young and intermediate-age star clusters. Gordon et al. (1999) note that the 
luminosity and spatial extent of the M33 nucleus is not greatly different from 
that of SgrA, where there is a well-documented young population. The current 
data indicate that the nuclei of M33 and the Galaxy, when viewed at the same 
spatial resolution, have comparable CO indices, providing 
yet another way in which the central regions of these galaxies are similar. 
Given the relativley low mass infered for the central black hole in M33 
(assuming that such an object is even present), then it appears that star 
formation in the central regions of spiral galaxies does not require the 
presence of a super massive compact object.

	The source of the gas that fueled the recent nuclear star formation 
in M33 is a matter of speculation. Kormendy \& McClure (1993) point 
out that it would be difficult for gas from the disk of M33 to find the 
nucleus given the shallow gravitational potential. However, Regan \& Vogel 
(1994) found a weak bar-like structure, which could serve as a conduit for 
channeling gas into the central regions of the galaxy.

	Aside from the blue nucleus, the central light concentration of M33 has 
a near-infrared SED similar to that of globular clusters. The metallicity of 
the central light concentration provides important 
clues into its origins, and the CO index between 
0.5 to 1.0 arcsec from the nucleus, where contamination from the central 
blue population should be small, is CO$_0 \sim 0.05$, which 
is indicative of a low metallicity. Davidge (1999) revised the relation 
between integrated CO strength in globular clusters and metallicity originally 
investigated by Aaronson et al. (1978), and this re-calibration predicts that 
[M/H] $\sim -1.2$ in the outer regions of the central light concentration, 
with an estimated uncertainty of 0.5 dex. While O'Connell (1983) and Schmidt et 
al. (1990) both conclude that a large metal-rich component is present near the 
center of M33, the data used in those investigations were recorded through 
relatively large apertures, so there is significant contamination from the 
inner disk; consequently, the metallicities infered from these studies 
do not necessarily apply to the central light concentration. 
A low metallicity for the M33 central light concentration is perhaps not 
surprising given the empirical relation between chemical composition and bulge 
mass derived from spectroscopic studies of early-type spirals (e.g. Jablonka, 
Martin, \& Arimoto 1996). The metallicity of spheroidal systems 
depends on the local escape velocity (e.g. Franx \& Illingworth 1990, 
Davies, Sadler, \& Peletier 1993, Martinelli, Matteucci, \& Colafrancesco 
1998), and the modest velocity dispersion near the center of M33 indicates that 
the escape velocity in this region is lower than near the centers of larger 
spheroids, such as the bulge of M31; consequently, a significant difference 
in mean metallicity is to be expected. Finally, a low metallicity suggests that 
the central light concentration could not have formed from chemically evolved 
disk material.

	The metallicity computed in the preceding paragraph assumes 
that the central light concentration is a simple stellar system, 
and colors measured at visible wavelengths provide a means of 
testing this assumption. Kormendy \& McClure (1993) find that $B-R \sim 
1.35$ 1 arcsec from the nucleus, which corresponds to $(B-R)_0 = 0.8$ if 
$E(B-V) = 0.3$, while Lauer et al. (1998) find that $V-I = 1.00$ when r $\geq 
0.3$ arcsec, so that $(V-I)_0 \sim 0.7$. For comparison, globular clusters 
typically have $B-R$ between 0.85 and 1.45 and $V-I$ between 0.7 and 1.0 (Reed, 
Hesser, \& Shawl 1988). Hence, the visible colors of the central light 
concentration appear to be lower than those predicted for an old population, as 
expected if a young component, which could influnece the CO index, is present. 
However, the amount of contamination from a young population, especially at 
infrared wavelengths, may be modest. Moreover, the 
difference between the colors of the central light concentration and those 
of globular clusters is greatest for the Kormendy \& 
McClure (1993) data, which have a markedly 
lower angular resolution than the present observations, and 
light from the blue nucleus will influence the $B-R$ 
colors from these data at sub-arcsec radii. Instruments currently 
being developed for 8 metre telescopes will allow 
visible and near-infrared spectra with angular resolutions comparable to, 
and perhaps even surpassing, that of the current observations to be recorded 
of the center of M33, and data of this nature will provide a powerful 
means of decoupling age and metallicity in the central light concentration.

\vspace{0.3cm}
	Sincere thanks are extended to Sidney van den Bergh for commenting on 
an earlier draft of this paper, and to an anonymous referee, whose suggestions 
greatly improved the manuscript.

\clearpage

\begin{table*}
\begin{center}
\begin{tabular}{rrccc}
\tableline\tableline
x & y & $K$ & $J-K$ & $H-K$ \\
\tableline
   -496.9 &  -404.7 &  17.145 &   1.308 &   0.437 \\
   -491.3 &  -492.4 &  16.931 &   1.541 &   0.515 \\
   -476.4 &   177.6 &  17.099 &   1.448 &   0.395 \\
   -474.0 &   452.2 &  16.948 &   1.514 &   0.415 \\
   -466.4 &  -176.5 &  16.877 &   1.315 &   0.293 \\
   -457.7 &   405.7 &  16.017 &   2.586 &   1.005 \\
   -457.1 &  -460.6 &  16.362 &   1.423 &   0.365 \\
   -443.3 &   258.0 &  16.794 &   1.402 &   0.383 \\
   -440.2 &   416.7 &  15.032 &   1.258 &   0.334 \\
   -434.0 &   152.2 &  17.158 &   1.134 &   0.299 \\
   -433.2 &   116.3 &  17.432 &   0.983 &   0.237 \\
   -429.7 &  -464.3 &  16.178 &   1.545 &   0.639 \\
   -424.2 &   191.5 &  16.689 &   1.373 &   0.328 \\
   -377.7 &     9.1 &  16.625 &   1.733 &   0.544 \\
   -335.2 &  -206.1 &  16.048 &   1.900 &   0.561 \\
   -334.7 &   -59.8 &  17.475 &   1.825 &   0.457 \\
   -333.2 &   129.5 &  17.253 &   1.361 &   0.317 \\
   -329.0 &  -143.2 &  17.094 &   1.750 &   0.419 \\
   -322.0 &   293.0 &  17.069 &   1.365 &   0.392 \\
   -313.5 &   -49.6 &  17.103 &   1.419 &   0.389 \\
   -313.3 &   128.1 &  17.215 &   1.260 &   0.285 \\
   -298.7 &   432.3 &  16.342 &   1.561 &   0.552 \\
   -283.0 &   -63.8 &  17.438 &   1.313 &   0.414 \\
   -282.4 &   -24.6 &  16.618 &   1.327 &   0.372 \\
   -247.2 &   256.4 &  17.019 &   1.483 &   0.521 \\
   -235.2 &   415.7 &  15.849 &   1.338 &   0.355 \\
   -231.0 &    -8.6 &  17.364 &   1.196 &   0.289 \\
   -219.2 &   420.9 &  16.230 &   1.405 &   0.391 \\
   -209.0 &   296.1 &  17.357 &   1.374 &   0.414 \\
   -207.3 &  -427.7 &  17.110 &   1.260 &   0.438 \\
   -190.2 &    93.4 &  16.993 &   1.446 &   0.367 \\
   -164.1 &   417.1 &  16.687 &   1.439 &   0.397 \\
   -155.8 &  -309.6 &  16.705 &   1.400 &   0.312 \\
   -130.2 &  -115.3 &  17.181 &   1.399 &   0.393 \\
   -125.6 &  -367.2 &  17.395 &   1.333 &   0.449 \\
   -125.5 &   240.4 &  17.305 &   1.079 &   0.237 \\
   -118.2 &   438.8 &  16.185 &   1.644 &   0.506 \\
    -90.6 &   -54.9 &  16.581 &   1.359 &   0.305 \\
    -85.6 &  -425.4 &  17.354 &   1.631 &   0.419 \\
    -69.3 &  -167.4 &  16.601 &   1.607 &   0.529 \\
    -68.8 &   245.5 &  16.500 &   1.424 &   0.430 \\
    -55.7 &  -244.3 &  16.948 &   1.267 &   0.360 \\
    -30.2 &  -457.7 &  17.411 &   1.316 &   0.326 \\
    -25.2 &    97.8 &  16.733 &   1.342 &   0.207 \\
    -11.9 &    71.2 &  16.649 &   1.457 &   0.401 \\
     10.4 &  -210.3 &  16.741 &   0.970 &   0.306 \\
     26.7 &  -289.8 &  16.375 &   1.521 &   0.395 \\
     28.3 &  -131.5 &  15.914 &   1.451 &   0.437 \\
     29.8 &    85.8 &  16.711 &   1.446 &   0.368 \\
     33.2 &  -415.4 &  16.376 &   1.461 &   0.354 \\
     41.6 &   117.2 &  17.567 &   1.373 &   0.337 \\
     45.6 &  -212.5 &  17.242 &   1.412 &   0.368 \\
     49.2 &    -7.9 &  17.234 &   1.432 &   0.262 \\
     56.3 &     2.9 &  17.262 &   1.313 &   0.284 \\
     57.8 &   160.1 &  16.401 &   1.490 &   0.438 \\
     70.8 &   208.5 &  16.003 &   1.658 &   0.553 \\
     85.8 &    30.3 &  16.525 &   1.458 &   0.365 \\
    111.2 &   207.0 &  16.512 &   1.423 &   0.373 \\
    118.9 &  -214.9 &  17.358 &   1.056 &   0.247 \\
    121.5 &  -459.8 &  17.264 &   1.356 &   0.251 \\
    123.5 &   307.0 &  16.638 &   1.239 &   0.255 \\
    131.9 &   259.1 &  17.326 &   1.562 &   0.411 \\
    135.5 &  -141.3 &  16.506 &   1.074 &   0.292 \\
    168.9 &    97.1 &  17.204 &   1.313 &   0.406 \\
    174.4 &  -320.2 &  16.840 &   1.434 &   0.403 \\
    189.4 &  -125.0 &  16.769 &   1.282 &   0.388 \\
    219.2 &   263.3 &  16.263 &   1.527 &   0.504 \\
    226.1 &  -404.6 &  16.721 &   1.296 &   0.324 \\
    231.1 &  -280.7 &  17.138 &   1.384 &   0.409 \\
    232.4 &   202.2 &  16.640 &   1.524 &   0.451 \\
    246.4 &  -470.5 &  17.368 &   1.341 &   0.336 \\
    246.4 &   -26.9 &  16.081 &   1.087 &   0.236 \\
    264.8 &   163.7 &  17.272 &   1.214 &   0.256 \\
    266.8 &  -316.9 &  16.578 &   1.356 &   0.365 \\
    269.8 &   274.4 &  16.723 &   1.020 &   0.164 \\
    296.6 &   184.5 &  16.967 &   1.450 &   0.371 \\
    297.3 &  -182.5 &  16.975 &   1.252 &   0.334 \\
    379.5 &   487.0 &  16.609 &   1.358 &   0.306 \\
    396.9 &     9.7 &  16.688 &   1.658 &   0.460 \\
    411.9 &  -396.8 &  16.398 &   1.308 &   0.260 \\
    474.8 &  -128.6 &  16.489 &   1.400 &   0.433 \\
\tableline
\end{tabular}
\end{center}
\caption{PHOTOMETRY OF STARS WITH $K \leq 17.5$ NEAR THE CENTER OF M33}
\end{table*}

\clearpage

\begin{table*}
\begin{center}
\begin{tabular}{lcccc}
\tableline\tableline
Galaxy & $J-K$ & CO & $(J-K)_0$ & CO$_0$ \\
\tableline
M33 & 0.85 & 0.25 & 0.7 & 0.26 \\
 & $\pm 0.03$ & $\pm 0.03$ & $\pm 0.06$ & $\pm 0.03$ \\
 & & & & \\
SgrA & 5.7 & --0.1 & 1.0 & 0.26 \\
 & $\pm 0.03$ & $\pm 0.03$ & $\pm 0.15$ & $\pm 0.03$ \\
\tableline
\end{tabular}
\end{center}
\caption{THE NEAR-INFRARED PROPERTIES OF THE NUCLEI OF M33 AND THE GALAXY}
\end{table*}

\clearpage

\parindent=0.0cm
\begin{center}
FIGURE CAPTIONS
\end{center}

\figcaption
[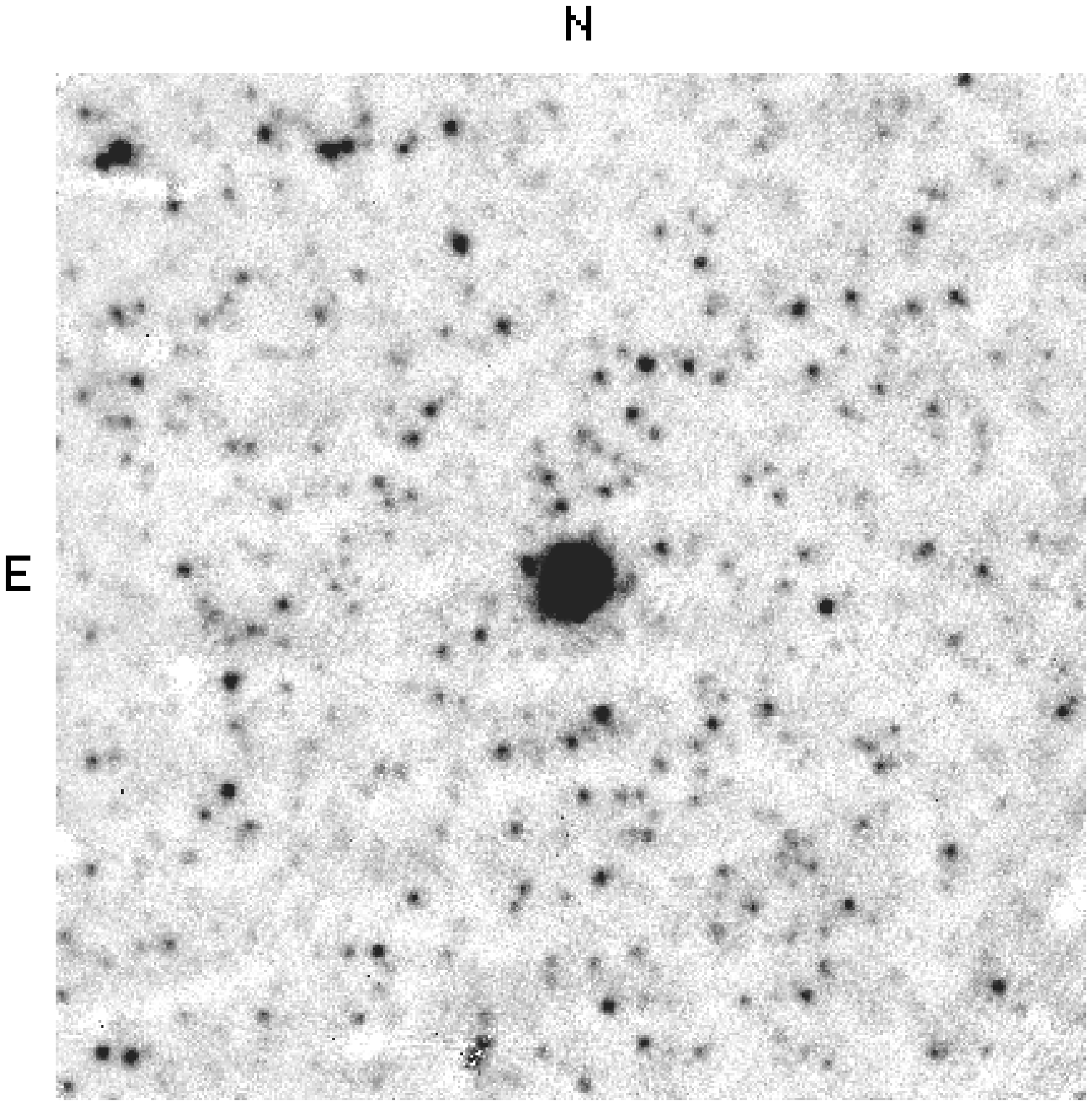]
{The final $K$ image of the M33 KIR field, which is roughly 
34 arcsec on a side. The central light concentration is the brightest 
source in the field.}

\figcaption
[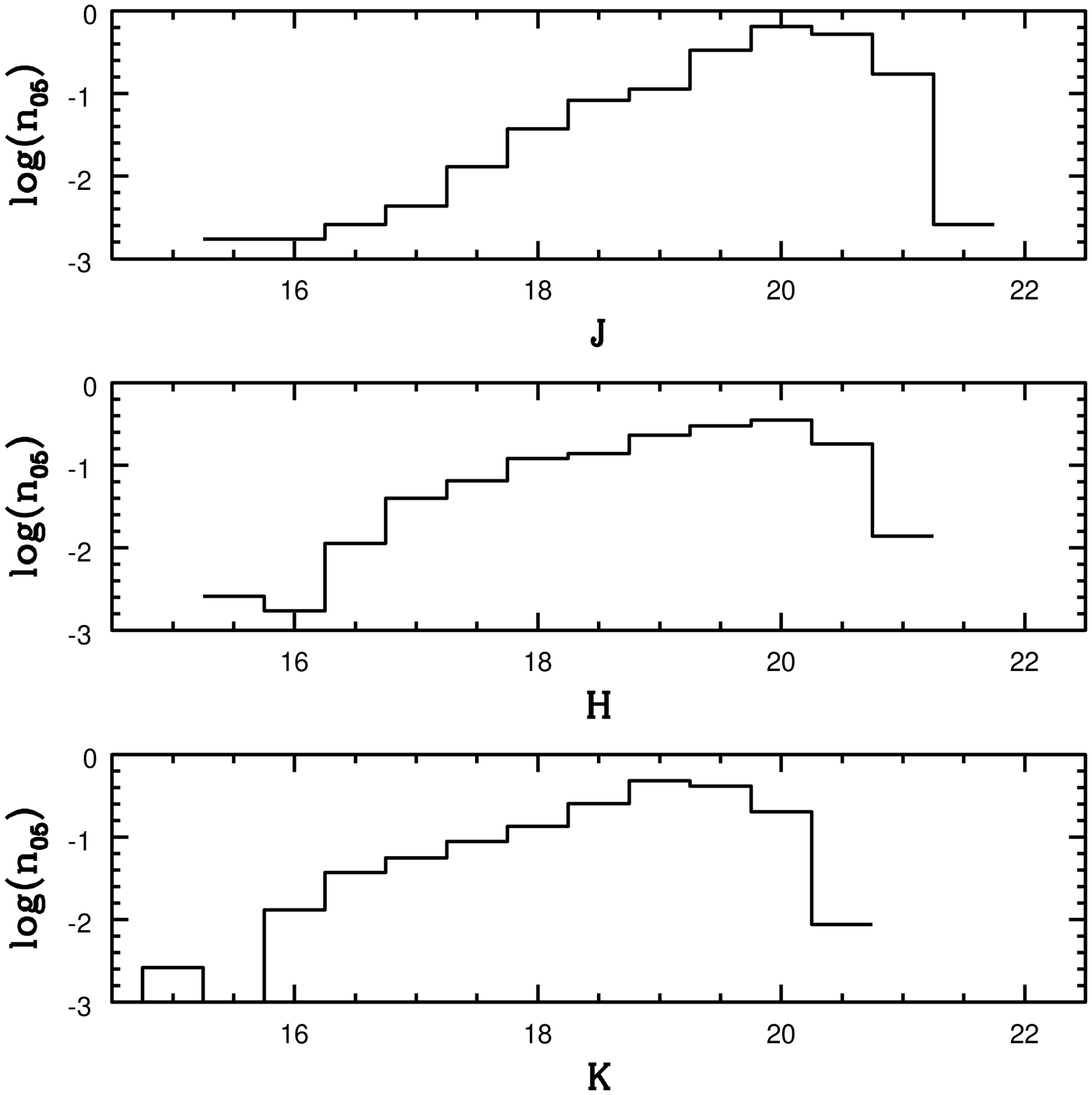]
{The $J, H,$ and $K$ LFs of the circumnuclear region. n$_{05}$ is the 
number of stars per square arcsec per 0.5 mag interval.}

\figcaption
[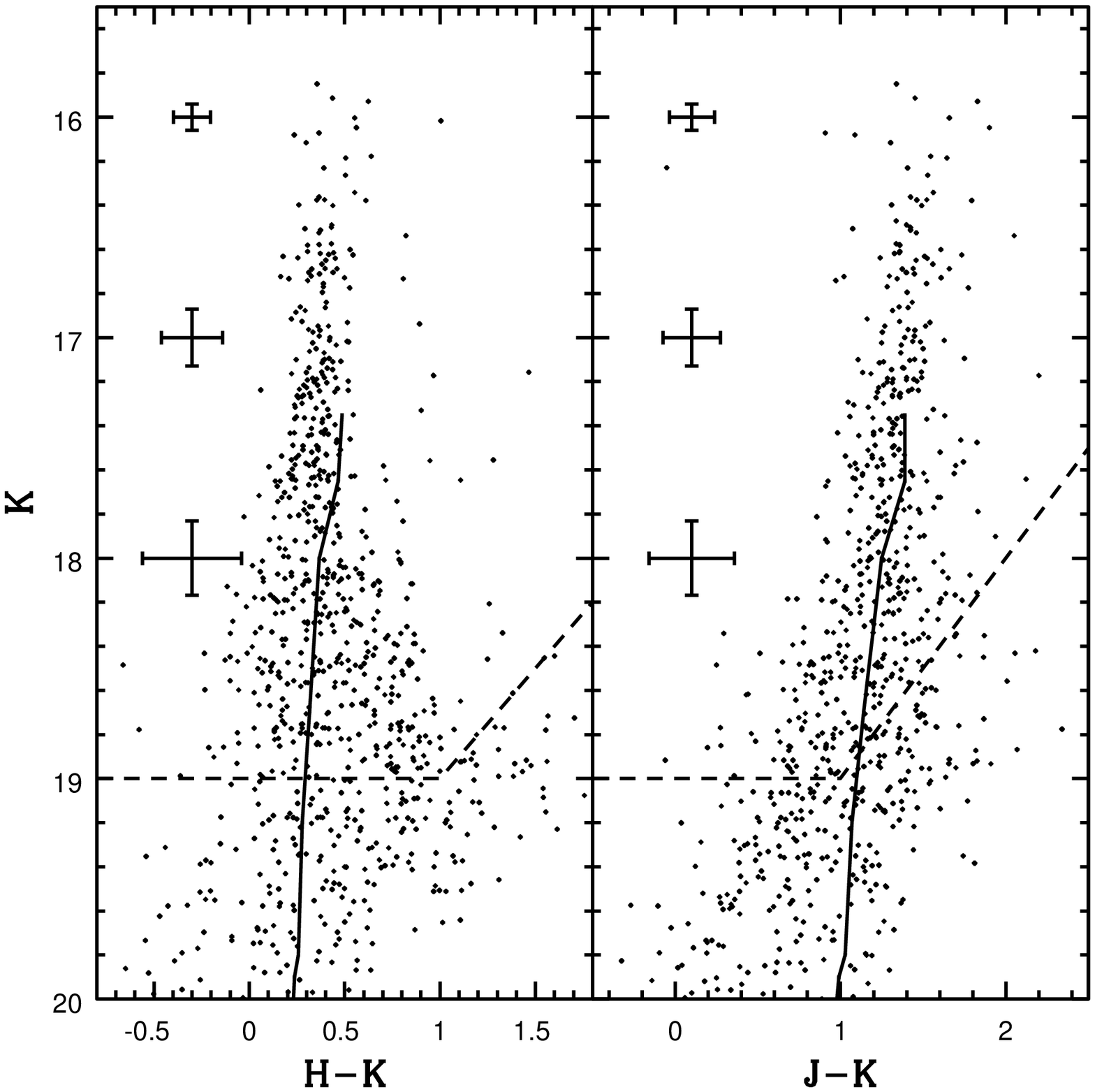]
{The $(K, H-K)$ and $(K, J-K)$ CMDs of the circumnuclear region. The dashed 
lines show the approximate completeness limits for these data, infered from the 
LFs, while the errorbars show the statistical uncertainties in the photometric 
measurements estimated from artificial star experiments. 
The solid line is the ridgeline of Baade's Window (BW) giants from 
Table 3B of Frogel \& Whitford (1987), shifted to a distance and reddening 
appropriate for the center of M33. A distance modulus of 24.5 
(van den Berg 1991) and E(B--V) = 0.3 (Wilson 1991) is assumed for M33, 
while a distance modulus of 14.5 (Reid 1993) is assumed for the Galactic 
Center.}

\figcaption
[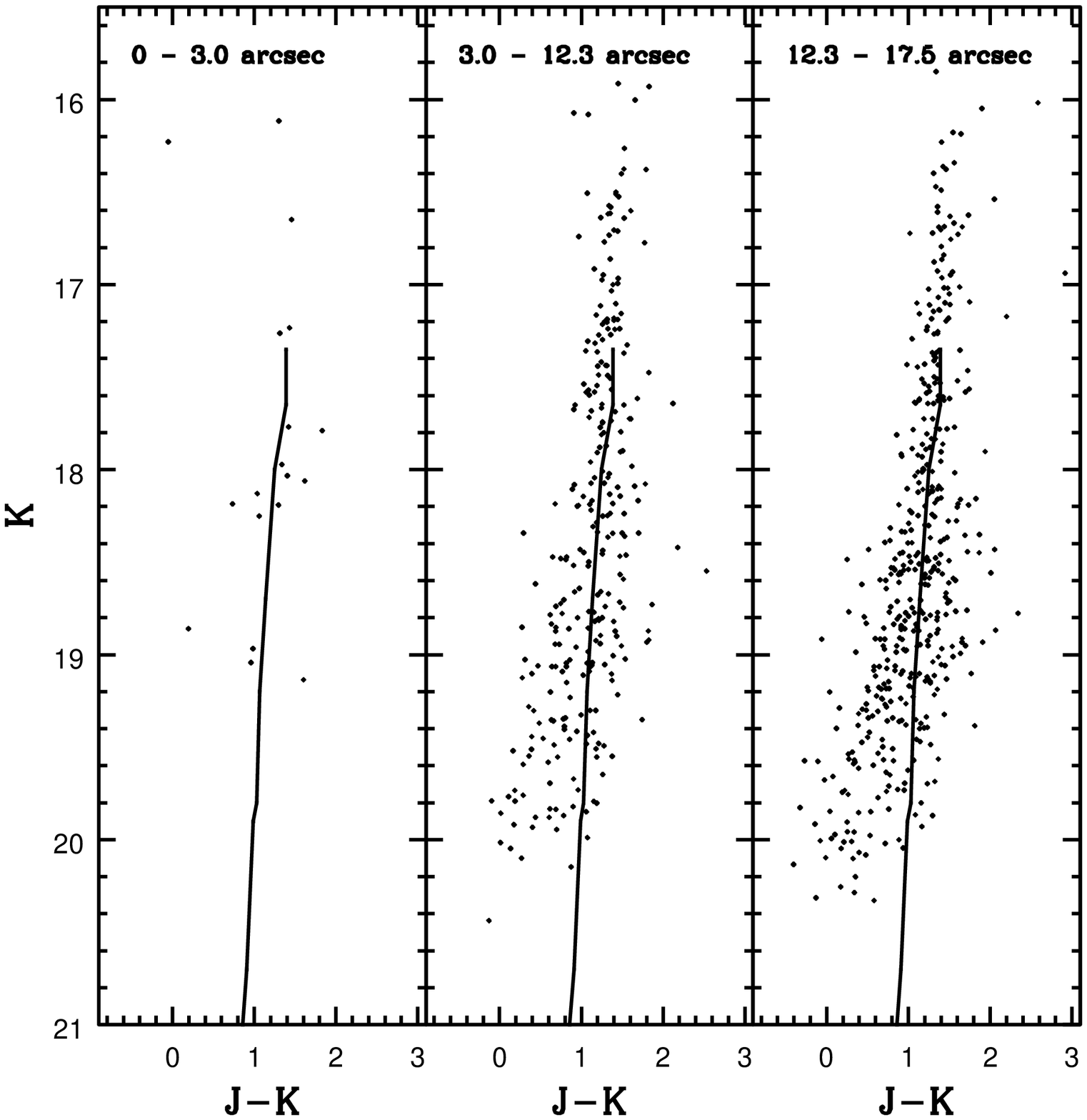]
{$(K, J-K)$ CMDs for three radial intervals, with distances 
measured from the center of M33. The solid line 
is the ridgeline of Baade's Window (BW) giants from 
Table 3B of Frogel \& Whitford (1987), shifted to a distance and reddening 
appropriate for the center of M33. A distance modulus of 24.5 
(van den Bergh 1991) and E(B--V) = 0.3 (Wilson 1991) has been assumed for M33, 
while a distance modulus of 14.5 (Reid 1993) is assumed for the Galactic 
Center.}

\figcaption
[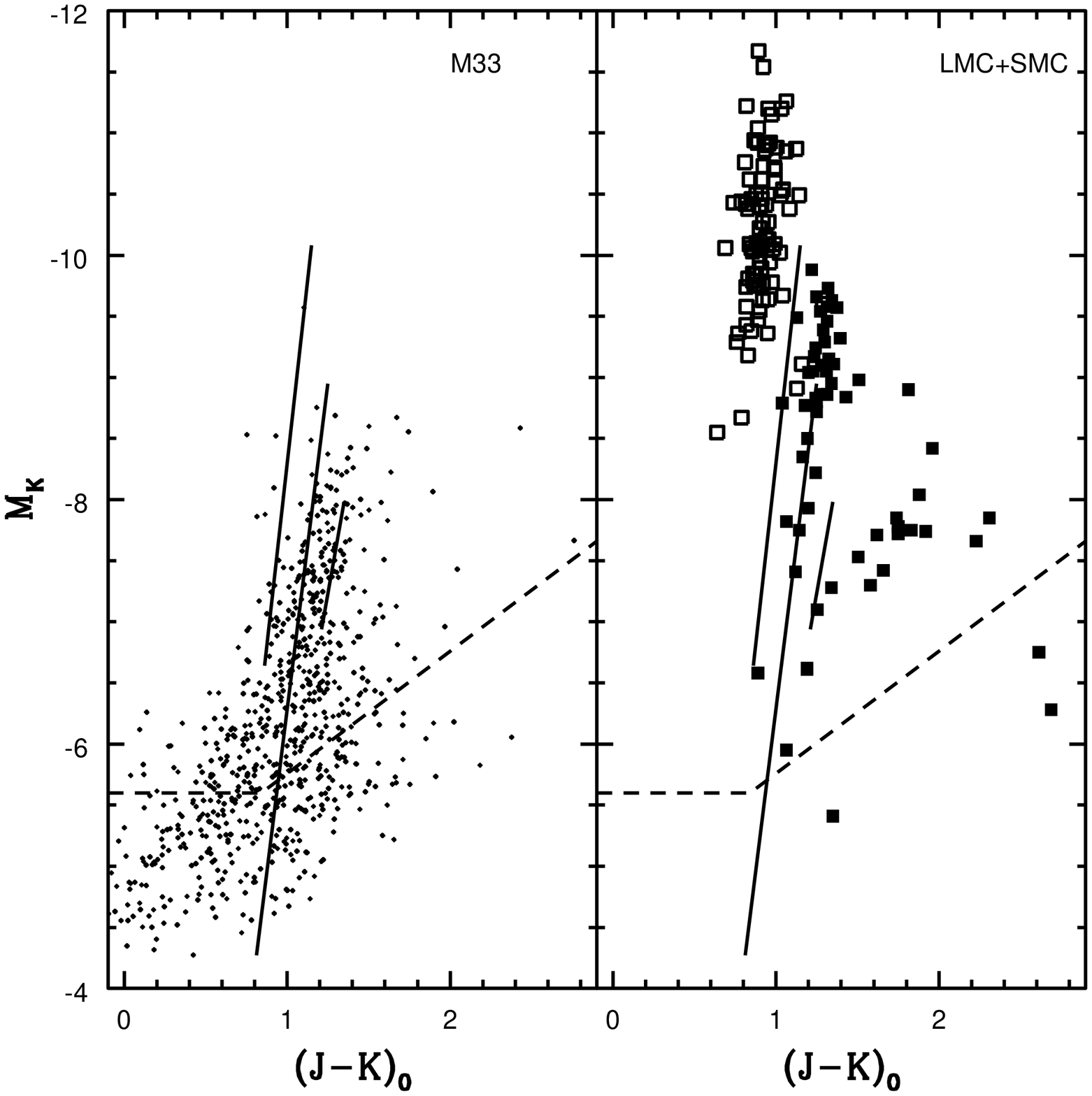]
{The (M$_K, (J-K)_0$) CMD for stars near the center of M33 is shown in the left 
hand panel, while the CMDs of RSGs (open squares) and LPVs (filled squares) 
in the Magellanic Clouds from data published by 
Elias et al. (1985) and Wood et al. (1983) are shown in the 
right hand panel. The dashed lines show the approximate completeness 
limits of the M33 observations, infered from 
the LFs in Figure 2. The solid lines are AGB sequences 
with ages of 0.1, 1.0, and 10 Gyr derived from the solar-metallicity 
Bertelli et al. (1994) models using the procedure described by Davidge (1998b).}

\figcaption
[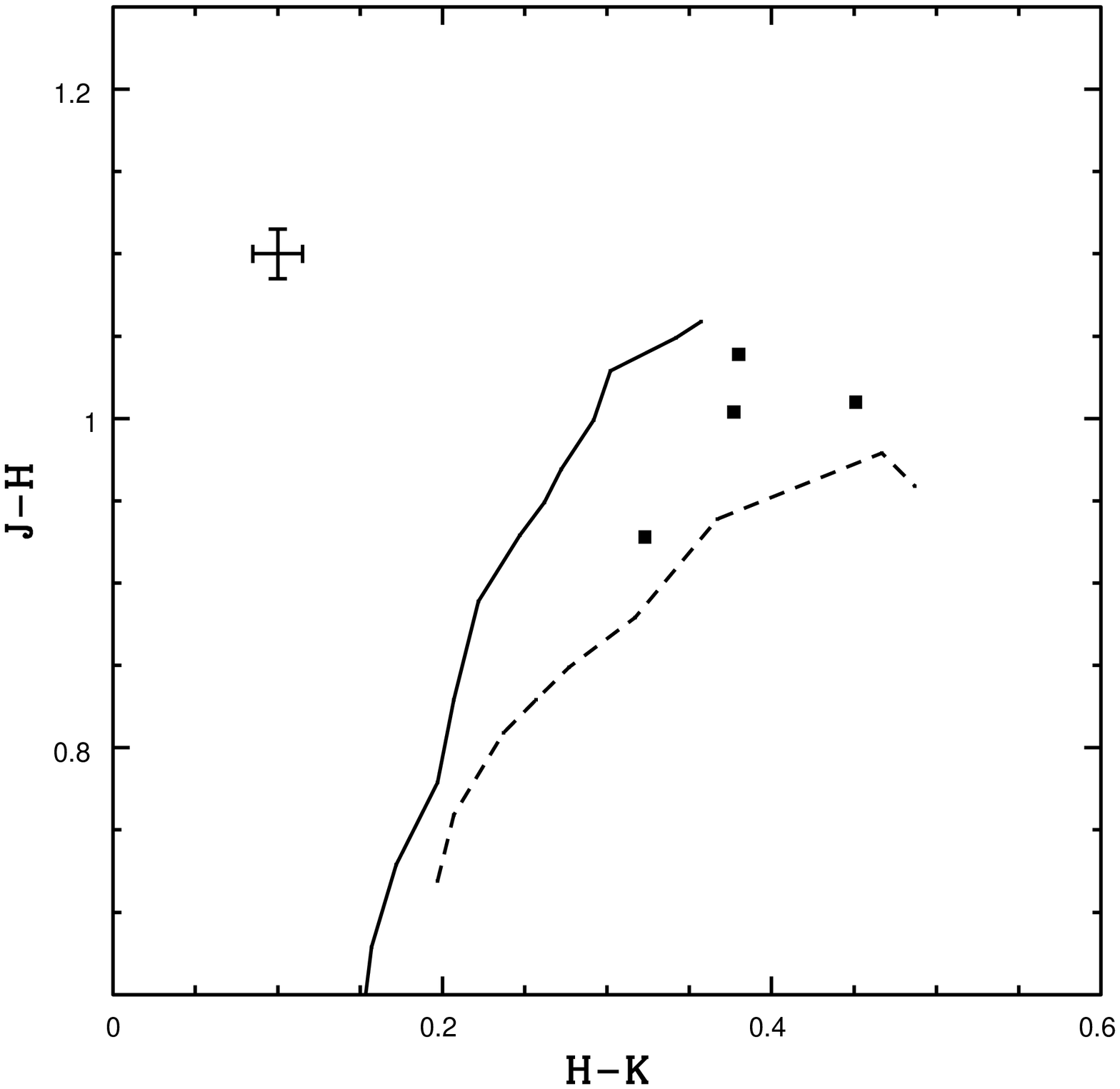]
{The $(H-K, J-H)$ TCD for stars near the center of M33. The filled squares 
show normal points computed in $\pm 0.25$ mag intervals along the $K$ axis of 
the $(K, H-K)$ and $(K, J-K)$ CMDs. The error bar in the upper left hand 
corner shows the random error in the normal point measurements. The solid 
line shows the sequence defined by solar neighborhood giants, 
while the dashed line shows the sequence for M giants in BW. Both of 
these sequences are from Frogel \& Whitford (1987), and 
have been reddened so that E(B--V) = 0.3.}

\figcaption
[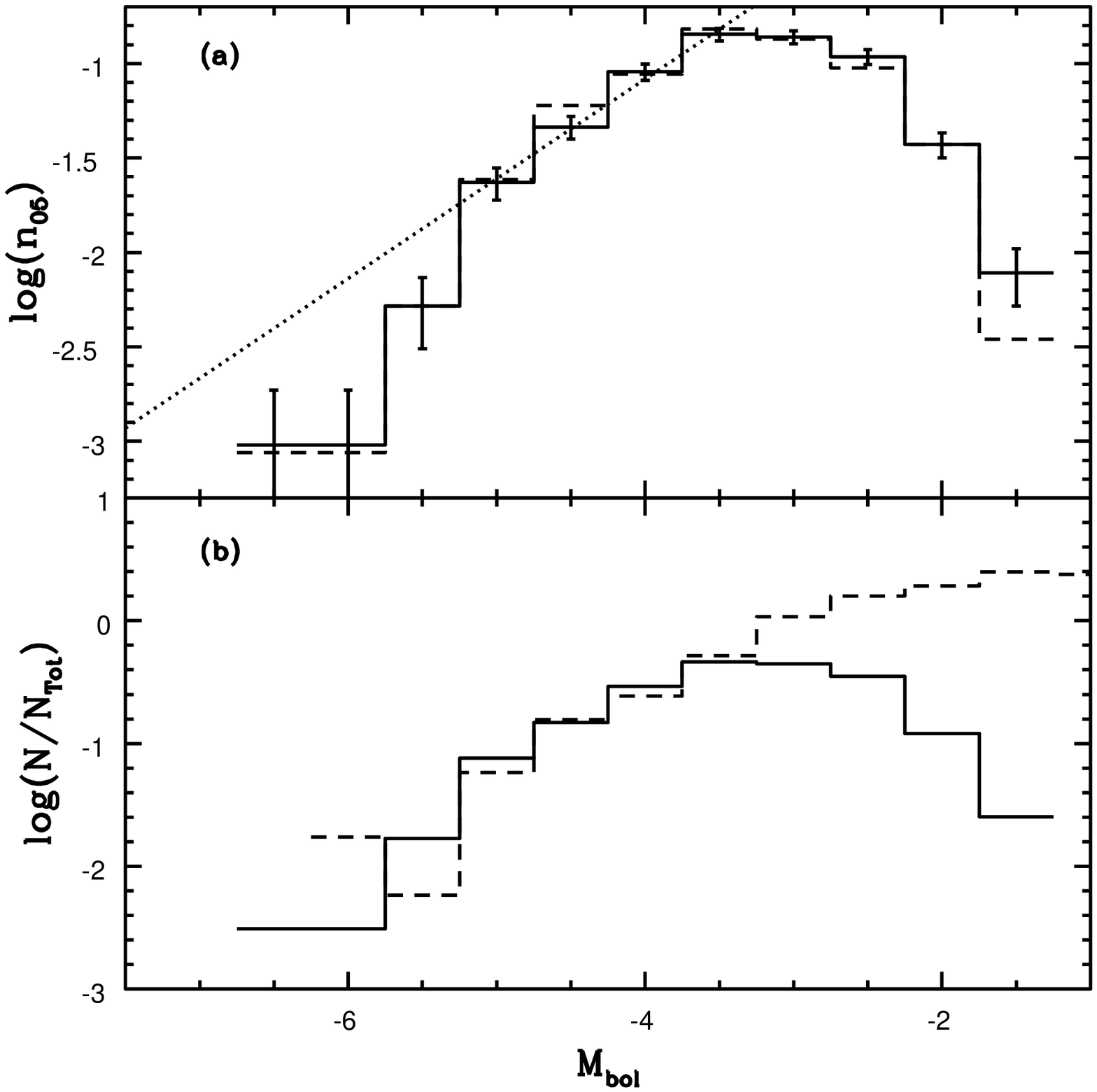]
{(a) Bolometric LFs for stars near the center of M33 derived with the 
bolometric correction calibrations for giants in BW (solid line) and the field 
(dashed line) from Figure 1b of Frogel \& Whitford (1987). n$_{05}$ is 
the number of stars per square arscec per 0.5 mag interval. The errorbars 
show the uncertainties in the LF computed with the BW bolometric correction 
calibration, while the dotted line shows a least squares fit to this 
LF in the interval -5 $\leq$M$_{bol} \leq -3.5$. (b) The bolometric 
LF for M33 derived from the BW bolometric correction calibration (solid line) 
compared with the LF for giants in the Galactic Center field studied by Davidge 
(1998a) (dashed line), which assumes that the distance modulus of the Galactic 
Center is $\mu_0 = 14.5$ (Reid 1993) and A$_K = 2.8$ (Davidge 1998a). 
N$_{Tot}$ is the number of stars with M$_{bol} \leq -3.25$.}

\figcaption
[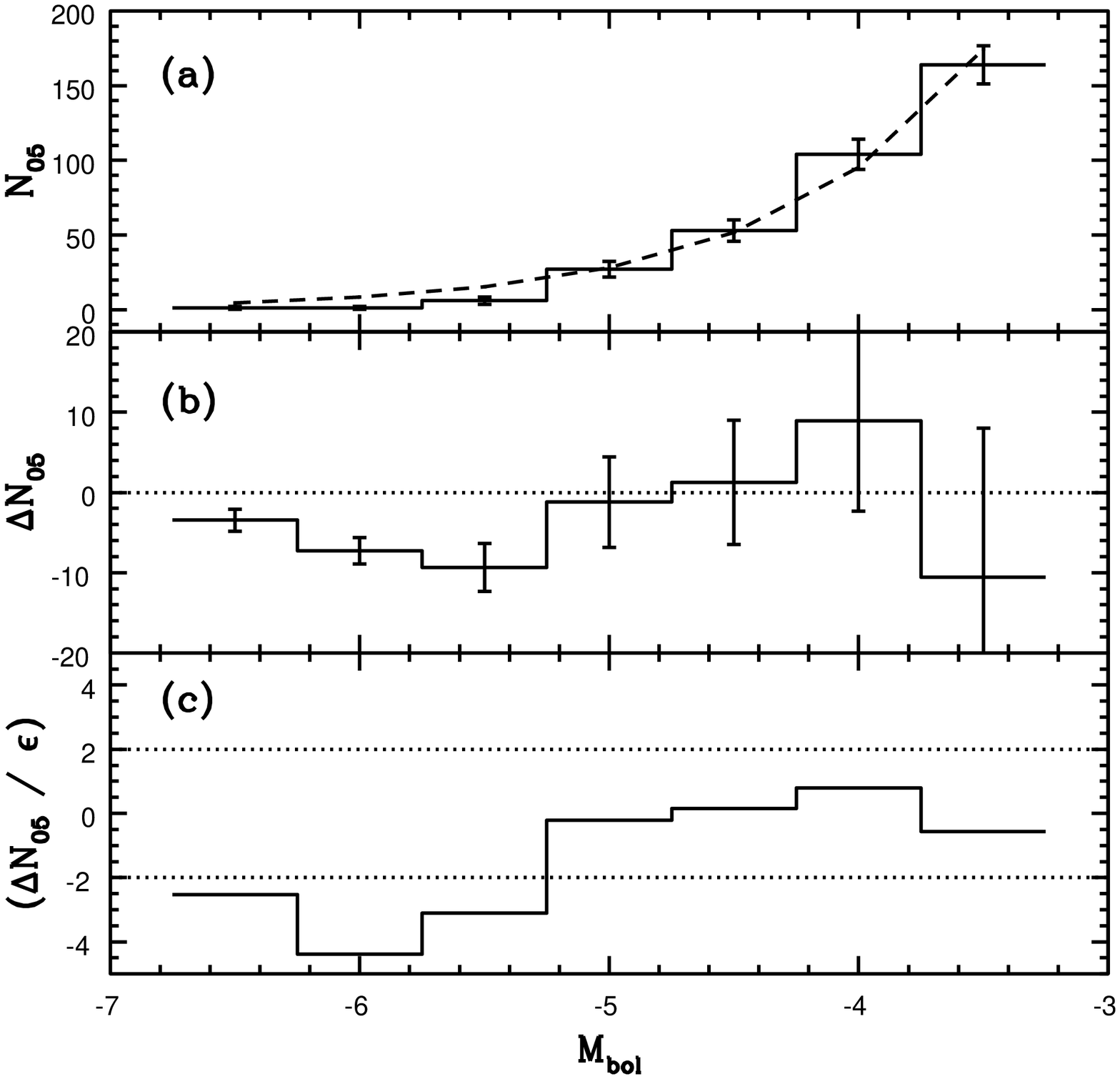]
{(a) The observed number counts in 0.5 mag bolometric luminosity bins, 
N$_{05}$, compared with the number counts predicted from the least squares fit 
shown in Figure 7 (dashed line). The error bars in this panel show the 
statistical uncertainties due to counting statistics. (b) The 
difference between the observed and predicted relations plotted in panel a. The 
error bars include not only the statistical uncertainies in each bin, but also 
the uncertainties in the least squares fit. (c) The result 
of dividing the $\Delta$N$_{05}$ entries in panel b by their 
estimated uncertainties. The dotted lines show the $\pm 2\sigma$ 
significance levels. The observations depart from the least squares fit 
in excess of the $2.5\sigma$ level when M$_{bol} \leq -5.25$, indicating that a 
break in the LF occurs near this brightness.}

\figcaption
[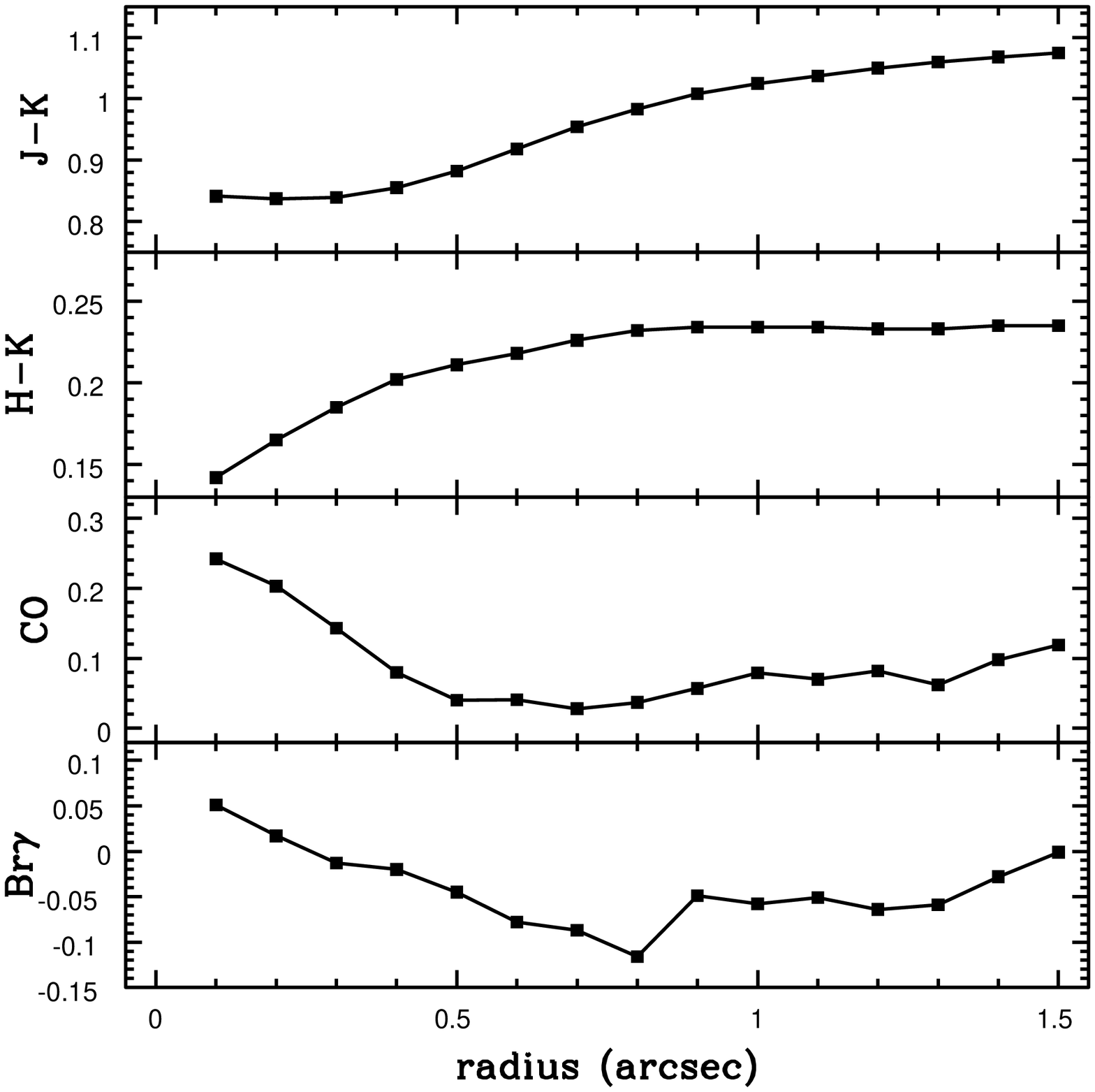]
{$J-K$, $H-K$, CO, and Br$\gamma$ profiles near the center of M33, 
with distance measured from the nucleus. The points plotted are averages within 
0.1 arcsec annuli. The Br$\gamma$ index is 
$-2.5 \times$ log(Br$\gamma$/$2.2\mu$m continuum), and the 
intensities shown for this index are in the instrumental system.} 

\figcaption
[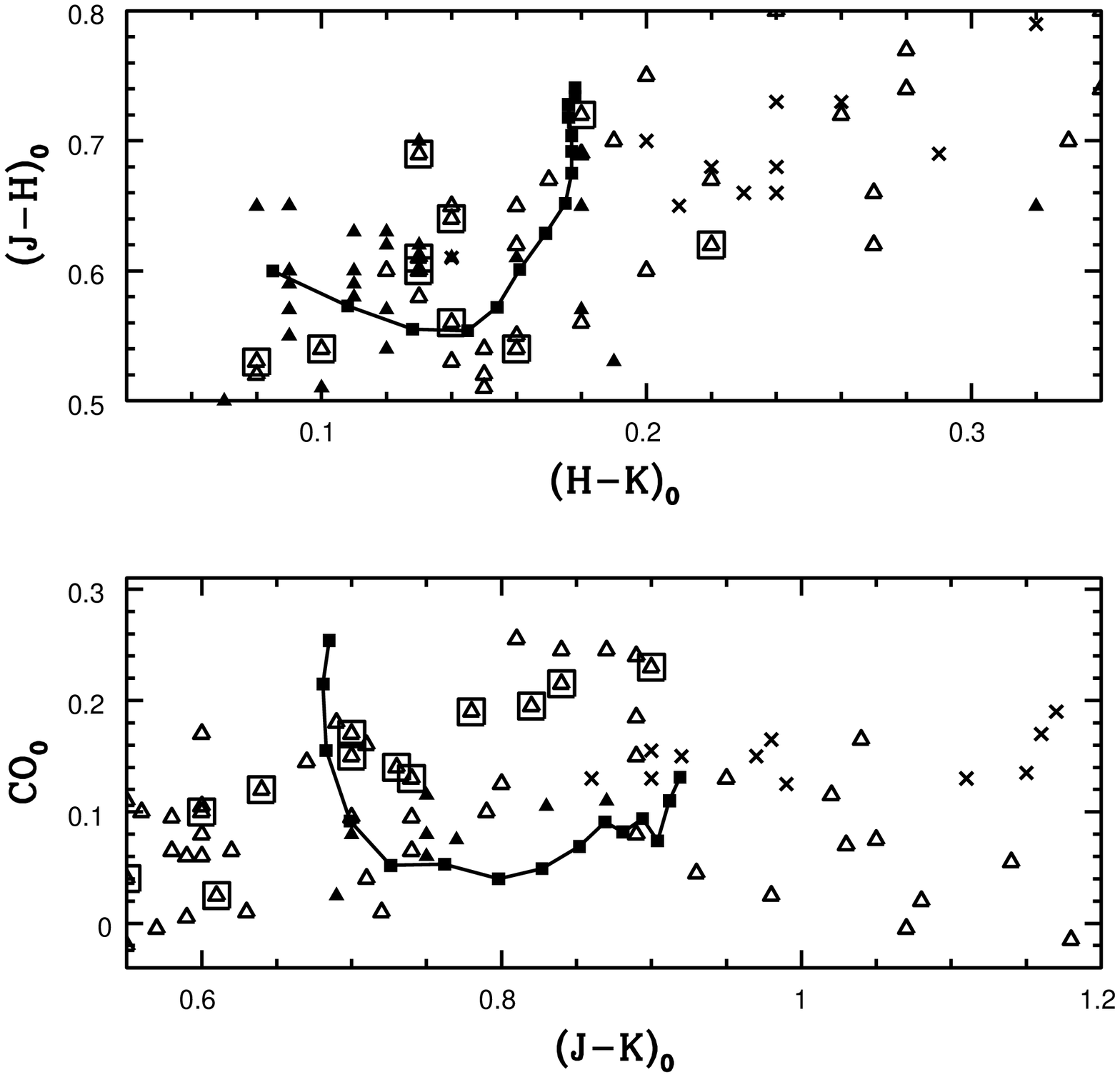]
{The $(J-H, H-K)$ and $(CO, J-K)$ TCDs of the central regions of M33. 
The M33 datapoints are averages within 0.1 arcsec annuli, and 
are shown as filled squares connected with the solid line; the progression 
from left to right for these data is one of increasing distance from the 
galaxy center. Also shown are data points for the central regions of 
nearby Sc galaxies (crosses; Frogel 1985), open clusters in the Magellanic 
Clouds (open triangles; Persson et al. 1983) and M31 globular clusters
(filled triangles; Frogel et al. 1980). Confirmed SWB type 1 and 2 
clusters, as identified in Tables 4 and 5 of Persson et al. (1983), are 
marked with a box; note that the four clusters with the highest CO indices 
do not have assigned SWB types.}

\figcaption
[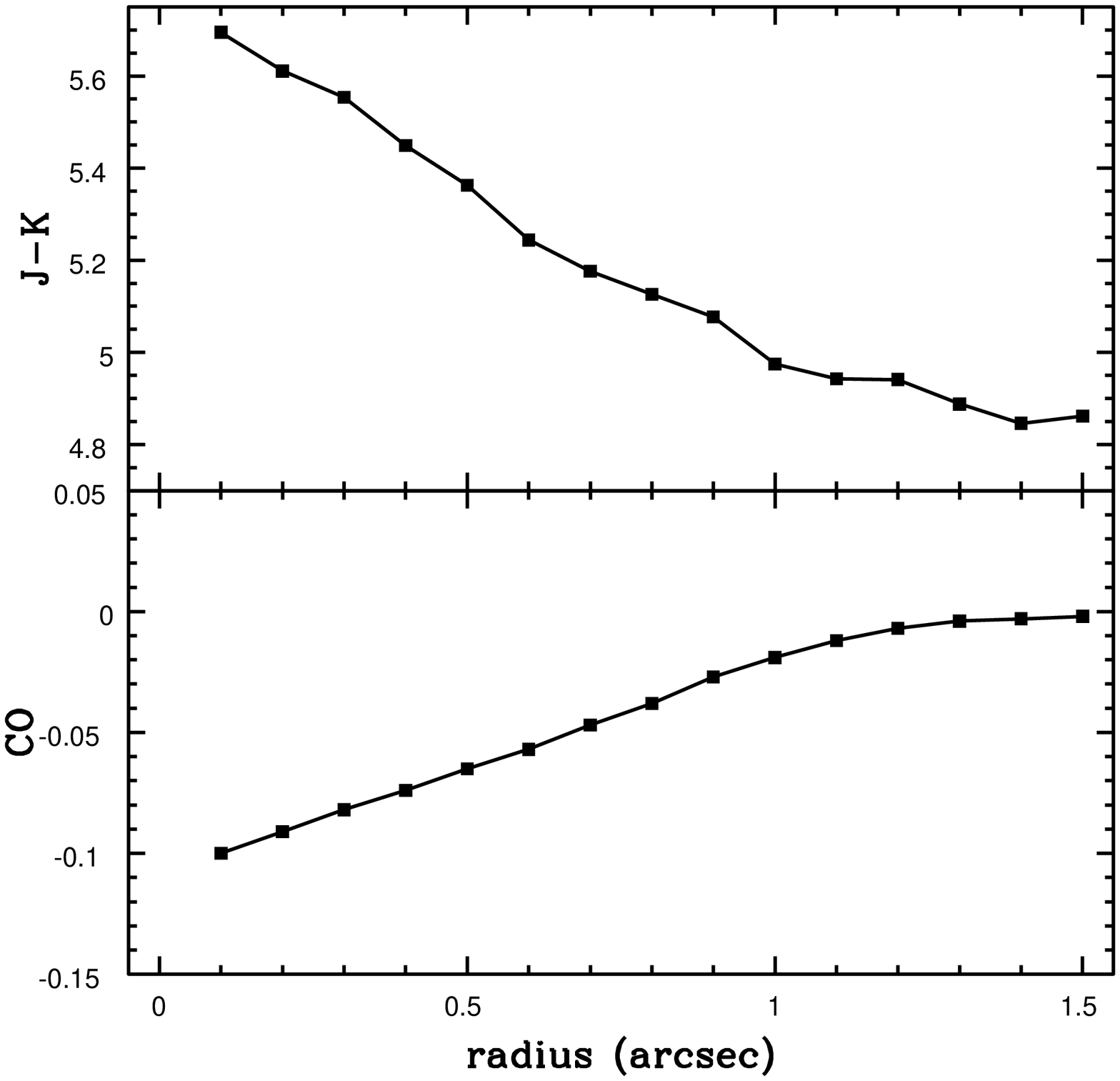]
{$J-K$ and CO profiles of the inner Galaxy, if viewed 
at the same distance as M33 with an angular resolution identical to that of 
the current observations. Averages in 0.1 arcsec annuli are plotted, 
and distances are measured from SgrA*. See text for additional details.}


\clearpage

\begin{references}

\reference{} Aaronson, M., Cohen, J. G., Mould, J., \& Malkan, M. 1978, ApJ, 
223, 824

\reference{} Andredakis, Y. C., Peletier, R. F., \& Balcells, M. 1995, 
MNRAS, 275, 874

\reference{} Barnes, J. E., \& Hernquist, L. E. 1992, ARA\&A, 30, 705

\reference{} Bertelli, G, Bressan, A., Chiosi, C., Fagotto, F., \& Nasi, E. 
1994, A\&AS, 106, 275

\reference{} Bothun 1992, AJ, 103, 104

\reference{} Casali, M., \& Hawarden, T. 1992, JCMT-UKIRT Newsletter, 4, 33

\reference{} Davidge, T. J. 1990, AJ, 99, 561

\reference{} Davidge, T. J. 1998a, AJ, 115, 2374

\reference{} Davidge, T. J. 1998b, ApJ, 497, 650

\reference{} Davidge, T. J. 1999, ApJS, in press

\reference{} Davidge, T. J., \& Courteau, S. 1999, AJ, 117, 1297

\reference{} Davidge, T. J., Le F\`{e}vre, O., \& Clark, C. C. 1991, ApJ, 
370, 559

\reference{} Davidge, T. J., Harris, W. E., Bridges, T. J., \& Hanes, D. A. 
1992, ApJS, 81, 251

\reference{} Davidge, T. J. et al. 1997, AJ, 114, 2586

\reference{} Davies, R. L., Sadler, E. M., \& Peletier, R. F. 1993, MNRAS, 
262, 650

\reference{} Dubus, G., Long, K. S., \& Charles, P. A. 1999, ApJ, 519, L135

\reference{} Elias, J. H., Frogel, J. A., \& Humphreys, R. M. 1985, ApJS, 57, 91

\reference{} Elias, J. H., Frogel, J. A., Matthews, K., \& Neugebauer, G. 1982, 
AJ, 87, 1029

\reference{} Franx, M., \& Illingworth, G. 1990, ApJ, 359, L41

\reference{} Frogel, J. A. 1985, ApJ, 298, 528

\reference{} Frogel, J. A., \& Whitford, A. E. 1987, ApJ, 320, 199

\reference{} Frogel, J. A., Mould, J., \& Blanco, V. M. 1990, ApJ, 352, 96

\reference{} Frogel, J. A., Persson, S. E., \& Cohen, J. G. 1980, ApJ, 240, 785

\reference{} Fullton, L. K. et al. 1995, AJ, 110, 652

\reference{} Garnett, D. R., Shields, G. A., Skillman, E. D., Sagan, S. P., 
\& Dufour, R. J. 1997, ApJ, 489, 63

\reference{} Gordon, K. D., Hanson, M. M., Clayton, G. C., Rieke, G. H., \& 
Misselt, K. A. 1999, ApJ, 519, 165

\reference{} Haller, J. W., Rieke, M. J., Rieke, G. H., Tamblyn, P., 
Close, L., \& Melia, F. 1996, ApJ, 456, 194

\reference{} Harris, W. E. 1996, 112, 1487

\reference{} Ho, L. C., Filippenko, A. V., \& Sargent, W. L. W. 1997, ApJ, 
487, 591

\reference{} Hodge, P. W. 1983, ApJ, 264, 470

\reference{} Jablonka, P., Martin, P., \& Arimoto, N. 1996, AJ, 112, 1415

\reference{} Kent, S. M. 1987, AJ, 94, 306

\reference{} Kormendy, J., \& McClure, R. D. 1993, AJ, 105, 1793

\reference{} Lauer, T. R., Faber, S. M., Ajhar, E. A., Grillmair, C. J., \& 
Scowen, P. A. 1998, AJ, 116, 2263

\reference{} Long, K. S., D'Odorico, S., Charles, P. A., \& Dopita, M. A. 
1981, ApJ, 246, L61

\reference{} Madau, P., Pozzetti, L., \& Dickinson, M. 1998, ApJ, 498, 106

\reference{} Martinelli, A., Matteucci, F., \& Colafrancesco, S. 1998, MNRAS, 
298, 42

\reference{} Massey, P., Bianchi, L., Hutchings, J. B., \& Stecher, T. P. 
1996, ApJ, 469, 629

\reference{} Matthews, L. D. et al. 1999, AJ, 118, 208

\reference{} McLean, I. S., \& Liu, T. 1996, ApJ, 456, 499

\reference{} Mighell, K. J., \& Rich, R. M. 1995, AJ, 110, 1649

\reference{} Minniti, D. 1995, AJ, 109, 1663

\reference{} Minniti, D., Olszewski, E. W., \& Rieke, M. 1993, ApJ, 410, L79

Morris, M., \& Serabyn, E. 1996, AAR\&A, 34, 645

\reference{} Mould, J. R., \& Kristian, J. 1986, ApJ, 305, 591

\reference{} Norman, C. A., Sellwood, J. A., \& Hasan, H. 1996, ApJ, 462, 114

\reference{} O'Connell, R. W. 1983, ApJ, 267, 80

\reference{} Ortolani, S. et al. 1995, Nature, 377, 701

\reference{} Persson, S. E., Aaronson, M., Cohen, J. G., Frogel, J. A., \& 
Matthews, K. ApJ, 266, 105

\reference{} Phillips, A. C., Illingworth, G. D., MacKenty, J. W., \& Franx, 
M. 1996, AJ, 111, 1566

\reference{} Reed, B. C., Hesser, J. E., \& Shawl, S. J. 1988, PASP, 100, 545

\reference{} Regan, M. W., \& Vogel, S. N. 1994, ApJ, 434, 536

\reference{} Reid, M. 1993, ARA\&A, 31, 345

\reference{} Rubin, V. C., \& Ford, W. K. Jr. 1986, ApJ, 305, L35

\reference{} Saha, P., Bicknell, G. V., \& McGregor, P. J. 1996, ApJ, 467, 636

\reference{} Schmidt, A. A., Bica, E., \& Alloin, D. 1990, MNRAS, 243, 620

\reference{} Searle, L., Wilkinson, A., \& Bagnuolo, W. G. 1980, ApJ, 239, 803

\reference{} Sellgren, K. McGinn, M. T., Becklin, E. E., \& Hall, D. N. B. 
1990, ApJ, 359, 112

\reference{} Shaver, P. A., McGee, R. X., Newton, L. M., Danks, A. C., \& 
Pottasch, S. R. 1983, MNRAS, 204, 53

\reference{} Stetson, P. B. 1987, PASP, 99, 191

\reference{} Stetson, P. B., \& Harris, W. E. 1988, AJ, 96, 909

\reference{} Trinchieri, G., Fabbiano, G., \& Peres, G. 1988, ApJ, 325, 531

\reference{} van den Bergh, S. 1976, ApJ, 203, 764

\reference{} van den Bergh, S. 1991, PASP, 103, 609

\reference{} Wilson, C. D. 1991, AJ, 101, 1663

\reference{} Wood, P. R., Bessell, M. S., \& Fox, M. W. 1983, ApJ, 272, 99

\end{references}
\end{document}